\documentclass[aps,twocolumn,showpacs,floatfix,amsfonts]{revtex4}

\usepackage{graphicx,graphics,color}
\usepackage{bm}
\usepackage{amsmath}
\usepackage{amssymb}
\usepackage{epsfig}

\newcommand{\beq}{\begin{equation}}
\newcommand{\eeq}{\end{equation}}
\newcommand{\beqarray}{\begin{eqnarray}}
\newcommand{\eeqarray}{\end{eqnarray}}
\newcommand{\half}{\ensuremath{\tfrac{1}{2}}}
\newcommand{\Hc}{\ensuremath{\mbox{H.c.}}} 
\newcommand{\Ham}[1][]{\ensuremath{{\cal{H}}_{\text{\tiny{#1}}}}} 
\newcommand{\eq}[1]{Eq.~(\ref{#1})} 
\newcommand{\fig}[1]{Fig.~\ref{#1}} 
\newcommand{\sgn}[1][]{\ensuremath{\text{sgn}(#1)}} 

\begin{document}

\title{Fractionally Charged Polarons and Phase Separation
in an Extended Falicov-Kimball Model}
\author{P. M. R. Brydon,$^1$ Jian-Xin Zhu,$^2$ and A. R. Bishop$^2$}
\affiliation{$^{1}$ Department of Theoretical Physics, Institute of
Advanced Studies,
The Australian National University, Canberra, ACT 0200, Australia\\
$^{2}$ Theoretical Division, Los Alamos National Laboratory, Los
Alamos, NM 87545, USA}

\date{\today}

\begin{abstract}
We present the first numerical study of the Falicov-Kimball model
extended by an on-site hybridization away from the particle-hole
symmetry point. Stable polaronic distortions of the charge-density
wave (CDW) phase are observed when doping with a single hole. For
moderate hole-doping, we find phase separation between a hole-rich
homogeneous state and the usual CDW state. Excitonic order is
enhanced around the hole-rich region, locally manifesting the global
competition between this and the CDW phase. Associated with the
hole-doping is a fractionalization
of the electron density between the itinerant and localized
electrons. The calculated local density of states at the polaron
centre reveals this fractionalization to be the result of charge
fluctuations induced by the hybridization. We propose a scanning
tunneling microscope experiment to test our results.
\end{abstract}

\pacs{71.28.+d, 71.45.Lr}
\maketitle

The Falicov-Kimball model (FKM) has found use
far beyond its
original role as a minimal model for valence transitions~\cite{FK69}.
The FKM describes itinerant $d$-electrons interacting via
a repulsive contact potential $U$ with localized $f$-electrons of
energy $\epsilon_{f}$: this simple physical picture is a useful
starting point for more complex models of mixed-valence (MV)
phenomenon~\cite{L78}, heavy fermions~\cite{epolarons},
electronic ferroelectricity~\cite{POS96,B02,YMDLH03} and
unconventional superconductivity~\cite{Miyake}. The ``bare'' form of
the FKM has also attracted much interest: it exhibits a charge density
wave (CDW) instability at half-filling in any dimension $D$ and is
considered a basic model of binary alloys~\cite{KL86}.

Apart from the extreme limits $D=1$~\cite{FGM96} and $D
\rightarrow\infty$~\cite{dinfty}, very little is known about the FKM
away from half-filling. The numerical results of Lema\'{n}ski
{\it{et al.}} suggest a rich 2D phase diagram, with phase separated
coexistence of many different orderings~\cite{LFB02}; it would
clearly be very interesting to assess the stability of these states
within the context of a more realistic model. The effect of dilute
hole doping on the CDW is also of interest: in related models
displaying density-wave instabilities (e.g. the Hubbard model)
hole-doping away from perfect nesting produces polarons~\cite{I97}.
Yin~\emph{et al.} have studied polarons in the
strong-coupling FKM with a weak $f$-electron hopping term, 
obtaining non-trivial modifications of the usual $t$-$J$ model
results~\cite{YMDLH03}. Of greater relevance, Liu and Ho have
proposed that polarons form in a FKM with hybridization, as a
precursor to a valence transition between states with integral and
fractional occupation of the $f$-orbitals~\cite{epolarons}.

In this Letter we extend our previous study of the FKM with $d$-$f$
hybridization away from the particle-hole symmetry
point~\cite{BZGB05}. We perform a Hartree-Fock (HF) decoupling of
the interaction and exactly diagonalize the resulting real-space
Bogoliubov-de Gennes (BdG) eigenequations.
We work exclusively in the limit of zero temperature where the HF
approach is known to be most accurate. Doping with a single hole
produces a local polaronic distortion of the CDW phase. Unique to
our model, we find a fractional partitioning of the electron density
between the two orbitals localized at the doped hole; this is
revealed by the local density of states (LDOS) spectra at the
polaron centre. We discuss the origin of the fractionalization and a
proposed experimental approach to observing this novel electronic
state. For a moderate hole-doping, we find that phase separation
between a hole-rich homogeneous phase and the usual half-filled CDW
phase is the ground state. The global competition between these
phases is manifested as a local enhancement of excitonic ordering.

We consider an extended FKM for spinless fermions:
\beqarray
\Ham & = & -\sum_{i,j}t_{ij}d^{\dagger}_{i}d_{j}
-\sum_{i,j}t^{f}_{ij}f^{\dagger}_{i}f_{j} +
\epsilon_{f}\sum_{j}n_{j}^{f} \notag \\
& & + V\sum_{j}\left\{d^{\dagger}_{j}f_{j}+\Hc\right\} +
  U\sum_{j}n^{d}_{j}n^{f}_{j}\;. \label{eq:HAM}
\eeqarray In addition to the usual FKM terms, we include a hopping
$t^{f}_{ij}$ for the $f$-electrons and an on-site hybridization $V$
between the $d$- and $f$-orbitals~\cite{note}. Here we concentrate
on the case $V\neq{0}$ and $t^{f}_{ij}=0$; we consider also a finite
$t^{f}_{ij}$ with $V=0$ for comparison with previous
work~\cite{B02,YMDLH03}. The hopping integrals $t_{ij}$ and
$t^{f}_{ij}$ are, respectively, $t$ and $t^{f}$ for nearest
neighbours, vanishing otherwise.

We perform the standard HF decomposition~\cite{L78} of the Coulomb
repulsion term: $n^{d}_{j}n^{f}_{j} =
\langle{n^{f}_{j}}\rangle{n^{d}_{j}} +
\langle{n^{d}_{j}}\rangle{n^{f}_{j}} -
\Delta_{j}d^{\dagger}_{j}f_{j} -
\Delta^{\ast}_{j}f^{\dagger}_{j}d_{j}$. Here $\Delta_{j} =
\langle{f^{\dagger}_{j}d_{j}}\rangle$ is the excitonic average at
site $j$. In the absence of a hybridization potential $V$,
$\Delta_{j}\neq0$ indicates the excitonic insulator
phase~\cite{KK65}; Portengen \emph{et al.} interpreted such a
``spontaneous'' excitonic average as evidence of electronic
ferroelectricity~\cite{POS96}. The resulting HF Hamiltonian is
diagonalized by the canonical transform,
$\gamma_{n}=\sum_{j}(u^{n}_{j}d_{j}+v^{n}_{j}f_{j})$. The
quasiparticle wavefunction amplitudes, $u^{n}_{j}$ and $v^{n}_{j}$,
are derived by solving the associated BdG eigenequations: \beq
\sum_{j}\left(\begin{array}{cc}
{\cal{H}}^{dd}_{ij} & {\cal{H}}^{df}_{ij} \\
{\cal{H}}^{df\ast}_{ji} & {\cal{H}}^{ff}_{ij}
\end{array}
\right)\left(\begin{array}{c}
u^{n}_{j} \\
v^{n}_{j}
\end{array}\right) =
E_{n}\left(\begin{array}{c}
u^{n}_{i} \\
v^{n}_{i}
\end{array}\right)\;, \label{eq:BdG}
\eeq where the components of the Hamiltonian matrix are defined as
${\cal{H}}^{dd}_{ij} = -t_{ij} +
U\langle{n^{f}_{j}}\rangle\delta_{ij}$, ${\cal{H}}^{ff}_{ij} =
-t^{f}_{ij} + (\epsilon_{f}+U\langle{n^{d}_{j}}\rangle)\delta_{ij}$
and ${\cal{H}}^{df}_{ij} = (V-U\Delta_{j})\delta_{ij}$. In terms of
the diagonal basis the order parameters at site $j$ are given by
$\langle{n^{d}_{j}}\rangle = \sum_{n=1}^{N_{tot}}|u^{n}_{j}|^{2}$,
$\langle{n^{f}_{j}}\rangle=\sum_{n=1}^{N_{tot}}|v^{n}_{j}|^{2}$ and
$\Delta_{j} = \sum_{n=1}^{N_{tot}}v^{n\ast}_{j}u^{n}_{j}$. We also
define the $d$- and $f$-electron CDW order parameters
$\delta^{d}_{j} = (-1)^{j}(\langle{n^{d}_{j}}\rangle-\half)$ and
$\delta^{f}_{j} = (-1)^{j}(\langle{n^{f}_{j}}\rangle-\half)$. Note
that in the CDW phase $\sgn[\delta^{d}_{j}] =
-\sgn[\delta^{f}_{j}]$. Our calculations are performed in the
canonical ensemble at zero temperature and thus the sum over $n$
extends over the first $N_{tot}$ occupied quasiparticle states.

We solve the BdG eigenequations self-consistently using a numerical
iteration scheme. Commencing with an initializing set of order
parameters, we exactly diagonalize~\eq{eq:BdG} and hence compute new
order parameters using the obtained quasiparticle wavefunctions.
These values are then used as an input for the next iteration; this
procedure is repeated until a desired accuracy is reached. The
calculations are performed for a $N=24\times24$ lattice with
periodic boundary conditions. To calculate the quasiparticle LDOS we
use the converged $24\times24$ lattice as a supercell in a
$10\times10$ array. The $d$-electron hopping integral $t$ defines
our energy scale. The results presented below correspond to
$\epsilon_{f}=0$ and $U=2.0$; for the case of a single hole we take
$V=0.1$ ($t^{f}=-0.1$), while $V=0.2$ ($t^{f}=-0.2$) for the case of
moderate hole-doping.

\begin{figure}[t]
\resizebox{\columnwidth}{!}{\includegraphics*[22mm,65mm][200mm,210mm]{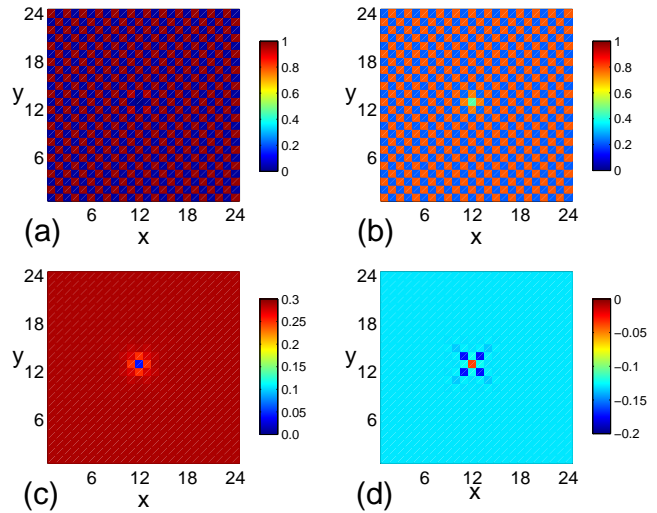}}
\caption{\label{pol1}(color online) The variation of the $f$-electron
  density (a), the $d$-electron density (b), the $d$-electron CDW
  order parameter (c), and the excitonic order parameter (d) in the
  vicinity of a single hole.}
\end{figure}

{\em Limit of a single hole.} We dope a single hole into the CDW
state by computing the order parameters for $N_{tot}=N-1$ (note
$N_{tot}=N$ corresponds to half filling); the converged results are
presented in~\fig{pol1}. The hole appears as a localized vacancy in
the occupied $f$-electron sub-lattice [see~\fig{pol1}(a)]. This
defect in the periodic HF potential $U\langle{n^{f}_{j}}\rangle$
experienced by the $d$-electrons produces a distortion of the
$d$-electron CDW state which extends over several lattice
constants~[\fig{pol1}(b,c)], defining the spatial extent of a
polaron. Near the polaron, the excitonic average $\Delta_{j}$
[\fig{pol1}(d)] shows a Friedel-like oscillation with its tails
extending along the diagonals. The enhancement of $|\Delta_{j}|$
about the $f$-hole indicates that the polaron in the FKM is a local
manifestation of the competition between the global excitonic
ordering and CDW states~\cite{BZGB05}.

Away from the polaron $\delta^{d}_{j}$ is uniform to within the
accuracy of the convergence, indicating the expected single
$d$-electron per unit cell in the CDW bulk. Overall, however, we
find a total $d$-electron population $N_{d}=288.1084$, as against
the case of $N_{d0}=288$ $d$-electrons in the usual half-filled CDW
state. A ``fraction'' $\Delta{N_d}=N_{d}-N_{d0}=0.1084$ of a
$d$-electron is associated with the polaron. Correspondingly, the
total $f$-electron population  $N_f=286.8916$ such that $\Delta
N_f=N_f - N_{f0}=-1.1084$, showing the transfer of charge from the
$f$- to the $d$-orbitals. Although the formation of a polaron about
the $f$-hole is indicated by Liu and Ho's work~\cite{epolarons},
their method did not allow
them to observe the fractionalization. Our result also differs from
the homogeneous distribution of fractional charge in the traditional
MV state~\cite{L78}: we find the fractional charge localized at the
polaron.

This fractional charge arises from the mixing of the orbital
wavefunctions by the hybridization. As a result of this mixing, the
number of
$d$- and $f$-electrons are not separately conserved. (Note that the
total number of electrons, $\hat{N}=\hat{N}_{d}+\hat{N}_{f}$,
remains constant.) The hybridization allows electrons in the
$d$-orbitals to tunnel into the $f$-orbitals and {\it{vice versa}}:
this implies charge fluctuations between the two orbitals leading to
the observed fractional populations. Although true also in the
half-filled CDW state, it is only with the introduction of the
inhomogeneities by hole-doping (i.e. the polarons) that the
fractional charge can be directly observed. Fractionally charged
polarons are therefore unique to multi-orbital models such as the
FKM considered here; there is no analogous effect in single-orbital
systems.

We find that the fractional  $d$-electron $\Delta{N_d}$ (or
$f$-electron $\Delta N_{f}$) is strongly dependent upon $U$ and $V$:
$\Delta{N_d}$ increases with both these parameters, although it
decreases as $V$ is raised above the critical value that
destabilizes the global CDW phase. According to our HF decomposition
the excitonic average enters into the equations as an effective
hybridization potential $-U\Delta_{j}$. In the CDW state, however,
this is only non-zero in the presence of a finite \emph{on-site}
hybridization~\cite{BZGB05}. As the magnitude of $\Delta_{j}$ is
directly proportional to both $U$ and $V$, the observed behaviour of
$\Delta{N_{d}}$ is easily understood.

\begin{figure}[t]
\includegraphics[width=\columnwidth]{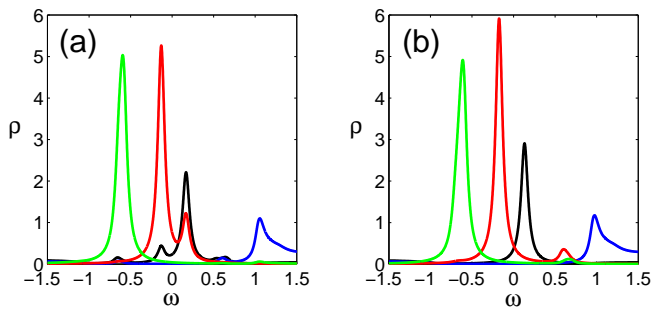}
\caption{\label{gapdos}(color online) The $d$- and $f$-electron LDOS
  spectra at the polaron centre and in the bulk for (a) $V=0.1$ and
  (b) $t^{f}=-0.1$ ($V=0$). In both panels the bulk $d$- and
  $f$-spectra are the blue and green lines, respectively; the polaron
  $d$- and $f$-spectra are given by the black and red lines,
  respectively.}
\end{figure}

The fractionalization can be explicitly visualized by computing the
LDOS both at the polaron site and in the CDW bulk. From our
converged solutions for the order parameters, we calculate the LDOS
of the $d$- and $f$-electrons at site $j$: \beqarray
\rho^{d}_{j}(\omega) &=&
\sum_{n}|u^{n}_{j}|^{2}\delta_{\alpha}(\omega_{n}-\omega),
\label{eq:LDOSd}
\\
\rho^{f}_{j}(\omega) & = &
\sum_{n}|v^{n}_{j}|^{2}\delta_{\alpha}(\omega_{n}-\omega).
\label{eq:LDOSf} \eeqarray Here $\delta_{\alpha}(\omega)$ is the
Lorentzian with phenomenological life-time broadening characterized
by the parameter $\alpha$. In our plots of the LDOS spectra we
always take $\alpha=0.05$. We plot the LDOS spectra
in~\fig{gapdos}(a); in the accompanying~\fig{gapdos}(b) we plot the
same spectra for the FKM extended by a finite $f$-hopping integral
$t^{f}=-0.1$.

The effect of the hybridization on the $d$- and $f$-wavefunctions
can be clearly appreciated in~\fig{gapdos}(a). The off-diagonal
terms ${\cal{H}}^{df}_{ij}$ in~\eq{eq:BdG} produce quasiparticles with
mixed nature, as evidenced by the coincidence of the peaks in the $d$-
and $f$-electron spectra. The situation is qualitatively different for
the FKM extended by a finite $f$-hopping integral
[\fig{gapdos}(b)]. Essentially a two-band Hubbard model, in this case
${\hat{N}}_{d}$ and ${\hat{N}}_{f}$ separately commute
with the Hamiltonian (i.e. ${\cal{H}}^{df}_{ij}=0$) and
the quasiparticles retain purely $d$- or
$f$-character. This can be seen in the polaron LDOS spectra as two
distinct peaks for the $d$- and $f$-electrons within the CDW
gap.

In spite of the orbital-mixing, the LDOS spectra of the two species
is still an experimentally accessible quantity; for example, we can
imagine a scanning tunneling microscope (STM) experiment where the
tunneling matrix only couples the tip to the $d$-orbitals. The
polaron could then be identified by examining the spatial
distribution of the $d$-electron LDOS at the energies corresponding
to the two intra-gap peaks in~\fig{gapdos}(a). This will distinguish
between the fractionally charged polaron and the usual Hubbard-like
polaron: in our system a peak in the $d$-electron LDOS will be
observed at both energies, as opposed to only a single peak at the
higher energy when $V=0$.

\begin{figure}[t]
\resizebox{\columnwidth}{!}{\includegraphics*[20mm,65mm][200mm,210mm]{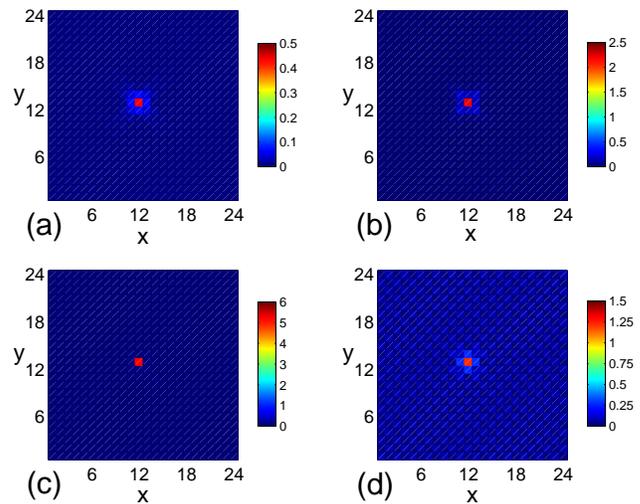}}
\caption{\label{DOS}(color online) The variation of the $d$-electron
  DOS $\rho^{d}(\omega)$ across the lattice for (a) $\omega=-0.131$
  and (b) $\omega=0.168$. In (c) and (d) we plot the $f$-electron DOS
  $\rho^{f}(\omega)$ for the energies corresponding to (a) and (b).}
\end{figure}

Using~\eq{eq:LDOSd} we calculate the variation of
$\rho^{d}_{j}(\omega)$ across the lattice for $\omega=-0.131$ and
$\omega=0.168$. The results are presented in~\fig{DOS}(a) and (b),
respectively; the corresponding LDOS for the $f$-electrons are given
in (c) and (d). As expected, for both species we find distinct peaks
in the LDOS at the polaron site for the two intra-gap energies. The
$d$-electron LDOS~\fig{DOS}(a) for the (primarily $f$-character)
$\omega=-0.131$ peak displays the diagonal star-shape seen before in
the plot of $\Delta_{j}$~[\fig{pol1}(d)]. Since the enhancement of
$|\Delta_{j}|$ near the polaron indicates strong $d$-$f$ mixing,
this structure in the LDOS is a defining signature of the
fractionally charged polaron. A much weaker diagonal structure is
present in the $f$-electron LDOS~\fig{DOS}(c). The same effect
occurs in the $\omega=0.168$ $d$-electron plot~[\fig{DOS}(b)], but
it is somewhat obscured by the higher $d$-electron peak; no such
structure is visible in the associated $f$-electron image.

\begin{figure}[]
\resizebox{\columnwidth}{!}{\includegraphics*[25mm,40mm][198mm,241mm]{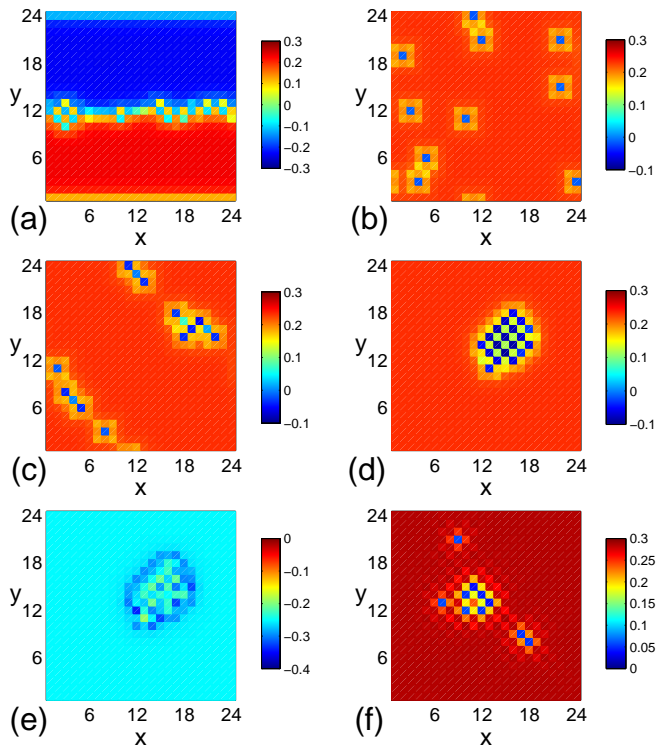}}
\caption{\label{tenhole}(color online) Variation of $\delta^{d}_{j}$
  for different initializing order parameter sets:
  (a) period-$12$ horizontal stripe; (b) CDW; (c) period-$24$
  diagonal stripe; (d) random. (e) Variation of $\Delta_{j}$ corresponding
  to (d). (f) Variation of $\delta_{j}^{d}$ for $t^{f}=-0.2$ ($V=0$) using same
  initializing sets as for (b).}
\end{figure}

{\em Finite hole doping.}  To study the effects of moderate
hole-doping, we solve the BdG eigenequations for $N_{tot}=N-10$. The
results for $V=0$ suggest a large variety of possible
charge-orderings~\cite{LFB02}; to find the ground state we therefore
consider a number of different initial order parameter sets. These
sets range from vertical and diagonal stripe orderings to a random
distribution of quasiparticle weight across the lattice. Since we
work in the zero temperature limit, we test the relative stability
of the converged results by calculating the total energy
$E=\sum_{n}E_{n}-U\sum_{j}(\langle{n^{d}_{j}}
\rangle\langle{n^{f}_{j}}\rangle-|\Delta_{j}|^{2})$. We present the
variations of $\delta^{d}_{j}$ for a representative sample of these
initializing sets in~\fig{tenhole}(a-d); these figures are labelled
with decreasing energy from $E/N=-1.5554$ for~\fig{tenhole}(a) to
$E/N=-1.5586$ for~\fig{tenhole}(d).~\fig{tenhole}(e) gives the
variation of $\Delta_{j}$ for the lowest-energy solution.
In~\fig{tenhole}(f) we present $\delta^{d}_{j}$ for finite $t^{f}$.

Our results clearly demonstrate the tendency for the $f$-holes to
cluster. Such phase separation is well known in the $V=0$ FKM, being
reported in the 2D model at all coupling
strengths~\cite{LFB02,FLU02}. Our observation of $f$-hole clustering
again confirms the relevance of the $V=0$ results as a guide to the
physics of the extended model~\cite{BZGB05}. An attractive
interaction between $f$-holes in a closely-related FKM has recently
been cited as a possible mechanism for the unconventional ``valence
fluctuation superconductivity'' observed in several Ce
compounds~\cite{Miyake}. Although our results cannot confirm such a
mechanism, they lend support to this scenario. With the inclusion of
a finite $f$-hopping, we see from~\fig{tenhole}(f) that there
appears to be some tendency for the phase separation to transform
into a diagonal stripe phase. This tendency can be understood by the
observation that the spinless FKM model with $t^{f}=t$ can be mapped
onto the Hubbard model, where the stripe phase is always obtained
upon finite hole doping~\cite{IMartin01}.

The fractionalization effect is present in~\fig{tenhole}(a-d); in
particular, for~\fig{tenhole}(d) we find $\Delta{N_{d}}=4.0038$,
indicating the transfer of four electrons from the $f$-orbitals. As
before, we find no evidence of this transfer in~\fig{tenhole}(f)
where $N_{d}=N_{d0}=288$. Also anticipated by our single-hole
results, we find enhancement of the excitonic order parameter
$|\Delta_{j}|$, mainly at the boundary of the $f$-hole cluster
[\fig{tenhole}(e)]. This again locally evidences the competition
between the excitonic and CDW orders.

In conclusion, we have investigated the FKM extended by the $d$-$f$
hybridization away from the half-filling symmetry point. Throughout
this report, we have compared and contrasted our results with the
extended FKM proposed in~\cite{B02}. We have studied the formation
of polarons when a single hole is doped into the CDW state.
Associated with the polaron, we have discovered a fractional
partitioning of the electron density between the $d$- and
$f$-orbitals. Such an effect is unique to our extended FKM: analysis
of the LDOS spectra at the polaron centre reveals the origins of the
fractionalization as the hybridization term. We propose a STM
experiment to verify our results. For moderate hole-doping, we have
demonstrated that phase separation into hole-rich homogeneous and
CDW states is energetically favorable. In both doping limits,
enhancement of the excitonic ordering is found around the hole-rich
region, locally manifesting the global competition between the two
phases~\cite{BZGB05}.

We thank W.-G. Yin for a careful reading of our manuscript and
constructive comments. The work at Los Alamos was supported by the
US DOE.


\begin{thebibliography}{99}
\bibitem{note}The on-site $d$-$f$ hybridization may be physically
  interpreted as an applied (optical) electric field.
\end{thebibliography}
\end{document}